\documentclass[a4paper,11pt]{article}
\usepackage{epsfig,amsmath,cite}

\newcommand\T{\rule{0pt}{2.6ex}}       
\newcommand\B{\rule[-1.2ex]{0pt}{0pt}} 

\newcommand{\be}{\begin{equation}}
\newcommand{\en}{\end{equation}}
\newcommand{\bea}{\begin{eqnarray}}
\newcommand{\ena}{\end{eqnarray}}
\newcommand{\lbl}[1]{\label{eq:#1}}
\newcommand{\lbltab}[1]{\label{tab:#1}}
\newcommand{\lblfig}[1]{\label{fig:#1}}
\newcommand{\lblsec}[1]{\label{sec:#1}}
\newcommand{\rf}[1]{(\ref{eq:#1})}
\newcommand{\Table}[1]{\ref{tab:#1}}
\newcommand{\fig}[1]{\ref{fig:#1}}
\newcommand{\sect}[1]{\ref{sec:#1}}
\newcommand{\gapprox}{%
\mathrel{%
\setbox0=\hbox{$>$}\raise0.6ex\copy0\kern-\wd0\lower0.65ex\hbox{$\sim$}}}
\newcommand{\lapprox}{%
\mathrel{%
\setbox0=\hbox{$<$}\raise0.6ex\copy0\kern-\wd0\lower0.65ex\hbox{$\sim$}}}
\setbox1=\hbox{$\scriptstyle\circ$}
\newcommand{\metapchir}{%
\mathrel{%
\setbox0=\hbox{$M$}\copy0\kern-0.6\wd0\raise1.02\ht0\copy1}}
\newcommand{\metapdchir}{%
\mathrel{
\setbox0=\hbox{$M$}
\hbox{$M^2_{\eta'}$}\kern-1.2\wd0\raise1.02\ht0\copy1}}

\newcommand{\mpid}{m_\pi^2}
\newcommand{\fpid}{F_\pi^2}

\newcommand{\mud}{\mu^2}

\newcommand{\piprim}{\pi'}
\newcommand{\cw}[1]{c^{W}_{#1}}
\newcommand{\cwr}[1]{c^{Wr}_{#1}}
\newcommand{\CW}[1]{C^{W}_{#1}}
\newcommand{\CWr}[1]{C^{Wr}_{#1}}
\newcommand{\Lr}[1]{L^r_{#1}}
\newcommand{\braque}[1]{{\langle #1 \rangle}}
\newcommand{\com}[1]{{[ #1 ]}}
\newcommand{\acom}[1]{{ \{ #1 \} }}
\newcommand{\ua}{u_\alpha}
\newcommand{\ub}{u_\beta}
\newcommand{\fpab}{ f_{+\alpha\beta} }
\newcommand{\fmab}{ f_{-\alpha\beta} }
\newcommand{\fpmn}{ f_{+\mu\nu} }
\newcommand{\fpgm}{ f_{+\gamma\mu} }
\newcommand{\fpga}{ f_{+\gamma\alpha} }
\newcommand{\fmmn}{ f_{-\mu\nu} }
\newcommand{\fmgn}{ f_{-\gamma\nu} }
\newcommand{\fmgb}{ f_{-\gamma\beta} }
\newcommand{\chip}{ \chi_+}
\newcommand{\chim}{ \chi_-}
\newcommand{\umu}{u_\mu}
\newcommand{\unu}{u_\nu}

\begin{document}
\hfill  PSI-PR-09-01\\[1cm]
\begin{center}
{\Large\bf Chiral expansions of the $\pi^0$ lifetime }
\bigskip

{\large K. Kampf\,$^{1,2}$ and B. Moussallam$^3$}

{\sl$^1$ Paul Scherrer Institut, Ch-5232 Villigen PSI, Switzerland}

{\sl$^2$ IPNP, Charles University, Faculty of Mathematics and Physics}\\
{\sl V Hole\v{s}ovi\v{c}k\'ach 2, CZ-180 00 Prague 8, Czech Republic}

{\sl$^3$\ Groupe de Physique Th\'eorique, Institut de Physique Nucl\'eaire}\\
{\sl Universit\'e Paris-Sud 11, F-91406 Orsay, France}

\end{center}
\vskip 1cm

\centerline{\bf Abstract}
The corrections induced by light quark masses to the
current algebra result for the $\pi^0$ lifetime are reexamined.
We consider next-to-next-to-leading order corrections and
we compute all the one-loop and the two-loop diagrams which contribute
to the decay amplitude at this order in the two-flavour chiral expansion.
We show that the result is renormalizable, as
Weinberg consistency conditions are satisfied. We find that
chiral logarithms are present at this order unlike the case at next-to-leading order.
The result could be used in conjunction with lattice QCD simulations, the
feasibility of which was recently demonstrated.
We discuss the matching between the two-flavour and the three-flavour
chiral expansions in the anomalous sector at order one-loop and derive the
relations between the coupling constants. A modified chiral counting
is proposed, in which $m_s$ counts as $O(p)$. We have updated
the various inputs needed and used this
to make a phenomenological prediction.

\section{Introduction}
The close agreement
between the current algebra prediction for the lifetime of the neutral
pion and experiment
is one of the two compelling experimental signatures, together with the
Nambu-Goldberger-Treiman relation,
for the spontaneous breaking of chiral symmetry in QCD.
There is an ongoing effort by the PrimEx collaboration~\cite{primex}
to improve significantly the accuracy of the lifetime measurement,
which is now around 8\%, down to the 1\%-2\% level.
This motivates us to study the corrections to the current
algebra prediction.

Starting with the detailed study by Kitazawa~\cite{kitazawa}, this problem
has been addressed several times in the
literature~\cite{donoghuelin,bijnenscornet,riazuddin,mou95,anantmou,nasrallah,kaisertempe,goityholstein,ioffe}.
The approach used in ref.~\cite{kitazawa}
was to extrapolate from the soft pion limit to the physical pion mass result
using the Pagels-Zepeda~\cite{pagelszep} sum rule method. This was reconsidered
in ref.~\cite{goityholstein} who implemented a more elaborate treatment
of $\pi^0-\eta-\eta'$ mixing and also recently in ref.~\cite{ioffe}.
The latter work shows some disagreement concerning the size of the $\eta'$
meson contribution in the sum rule as compared to earlier results.

In this paper, we reconsider the issue of the corrections to the current
algebra result to the $\pi^0\to 2\gamma$ amplitude from the point of view
of a strict expansion as a function of the light quark masses. This is
most easily implemented by using chiral Lagrangian methods
(see e.g.~\cite{scherrer02} for a review).
The same framework
also allows one to compute radiative corrections~\cite{urech95}.
We believe that
it is somewhat easier to control the size of the errors in this kind of
approach, which is important for  exploiting the forthcoming
high experimental accuracy. Another interest in deriving a quark mass
expansion is the ability to perform comparisons with lattice QCD
results where quark masses can be varied. The feasibility of
computing the $\pi^0$ to two photon amplitude in lattice QCD
has been studied very recently~\cite{cohen08}.

A priory, it is expected that one can make use of $SU(2)$ ChPT, i.e.
expand as a function of $m_u$, $m_d$ without making any assumption
concerning $m_s$ (except that it is heavier than $m_u$, $m_d$). In $SU(2)$ ChPT
it is often the case that chiral logarithms provide a reasonable order of
magnitude for the size of the chiral corrections.
This is the case, for instance, for the $\pi\pi$ scattering
amplitude~\cite{colangelo1,colangelo2}.
It was observed in refs.~\cite{donoghuelin,bijnenscornet}
that there was no chiral logarithm in the next-to-leading order (NLO) correction to
the $\pi^0$ lifetime once the amplitude is expressed in terms
of the physical value of $F_\pi$.
We have asked ourselves whether chiral
logarithms are present in the NNLO corrections. At this order, the
coefficient of the double chiral logarithm depends only on $F_\pi$.
Depending on its numerical coefficient,
such a term could modify the NLO results.
In order to obtain
this  coefficient it is, in principle,
sufficient to compute a set of one-loop graphs
containing one divergent NLO vertex~\cite{weinberg79}. For completeness, we
will perform a complete calculation of the two-loop graphs as well.
This is described in sec.~\sect{NNLOcalc}.

In the framework of two-flavour ChPT, however,
one faces the practical problem that the polynomial terms in
$m_u,\ m_d$ at NLO involve a number of low-energy couplings (LEC's)
which are not known.
We will show that it is possible to make estimates for the relevant
combinations,
and then make quantitative predictions for the $\pi^0$ decay, under the minimal
additional assumption that the mass of the strange quark is sufficiently
small, justifying a chiral expansion in $m_s$. We will obtain the first
two terms in the $m_s$ expansion of the NLO $SU(2)$ LEC's. The result can
be implemented in association with a modified chiral counting scheme, in
which $m_s$ is counted as $O(p)$, which respects the hierarchy
$m_u,\ m_d \ll m_s$. This leads to simpler formulas than previously obtained.
Finally, we will update all the inputs needed to compute the lifetime.

\section{Leading and next-to-leading orders in the $SU(2)$ expansion}\lblsec{SU2exp}
In the odd-intrinsic-parity sector,
the Lagrangian  of lowest chiral order has order $p^4$, it is the
Wess-Zumino~\cite{WZ} Lagrangian, ${\cal L}^{WZ}$,
which form is dictated by the ABJ anomaly~\cite{ABJ}.
Writing the $\pi^0\to \gamma(k_1) \gamma(k_2)$ decay amplitude in the form
\be
{\cal T}= e^2\epsilon(e_1^*,k_1,e_2^*,k_2) T\ ,
\en
a tree level computation of the pion decay amplitude gives the well
known result
\be\lbl{Alo}
T_{LO}= {1\over 4\pi^2 F}\ ,
\en
where $F$ is the pion decay constant in the two-flavour chiral
limit $m_u=m_d=0$. At leading order one can set $F=F_\pi$ in eq.~\rf{Alo}.
According to the Weinberg rules\cite{weinberg79} for ChPT,
the NLO corrections are generated from:
\begin{itemize}
\item[a)] One-loop diagrams with one vertex taken from  ${\cal L}^{WZ}$
and other vertices from the $O(p^2)$
chiral Lagrangian. These diagrams were first computed in
refs.~\cite{donoghuelin,bijnenscornet}.
\item[b)] Tree diagrams having one vertex from
${\cal L}^{WZ}$ and one vertex from the $O(p^4)$ chiral Lagrangian.
\item[c)]Tree diagrams from the $O(p^6)$ Lagrangian in the
anomalous-parity sector, ${\cal L}^{W}_{(6)}$.
\end{itemize}
The classification of a minimal set of independent
terms in this Lagrangian was initiated in refs.~\cite{fearing,akhoury}.
We will use here the result of ref.~\cite{girlanda} who
further reduced the set to 23 terms in the case of three flavours
and to 13 independent terms in the case of two flavours (this result
was also obtained in ref.~\cite{eberthauser}). The list,
in the case of two flavours, is recalled below:
\begin{alignat}{3}
{\cal L}^{W}_{6,N_f=2}= \epsilon^{\alpha\beta\mu\nu}\Big\{
\;&\cw{1}\braque{\chip\com{\fmmn,\ua\ub}}
&+&\cw{2}\braque{\chim\acom{\fpmn,\ua\ub} }
&+&\cw{3}i\braque{\chim\fpmn\fpab}\notag\\
+&\cw{4}i\braque{\chim\fmmn\fmab}
&+&\cw{5}i\braque{\chip\com{\fpmn,\fmab} }
&+&\cw{6}\braque{\fpmn}\braque{\chim\ua\ub }\notag\\
+&\cw{7}i\braque{\fpmn}\braque{\fpab\chim}
&+&\cw{8}i\braque{\fpmn}\braque{\fpab}\braque{\chim}
&+&\cw{9}i\braque{\fpgm}\braque{h^\gamma_{\:\nu}\ua\ub}\notag\\
+&\cw{10}i\braque{f^{\,\gamma}_{+\ \mu} }\braque{\fmgn\ua\ub }
&+&\cw{11}\braque{\fpmn}\braque{\fpga h^\gamma_{\:\beta}}
&+&\cw{12}\braque{\fpmn}\braque{f^{\,\gamma}_{+\ \alpha}\fmgb}\notag\\
+&\cw{13}\braque{\nabla^\gamma\fpgm}\braque{f_{+\nu\alpha}\ub}
& \Big\} \ .\lbl{lagwz6}
\end{alignat}
The relations between the bare and the renormalized couplings
may be written as~\cite{girlanda}
\be\lbl{divciw}
\cw{i} = c_i^{Wr}(\mu) + \eta_i^W {(c \mu)^{d-4} \over 16\pi^2(d-4)}
\en
with $\log(c)=-(\log(4\pi)-\gamma+1)/2$ as usual in ChPT
(note that the couplings $\cwr{i}$ have dimension $(mass)^{-2}$).
The coefficients $\eta_i^W$ vanish for $i=1...5$ and the remaining
ones read~\cite{girlanda}
\be\lbl{etadiv}
\begin{array}{llll}
   \eta_6^W= 3\alpha,\
 & \eta_7^W= 3\alpha,\
 & \eta_8^W= -{3\over2}\alpha,\
 & \eta_9^W= 6\alpha             \\[0.2cm]
   \eta_{10}^W= -18\alpha,\
&  \eta_{11}^W= 12\alpha,\
&  \eta_{12}^W= 0,\
&  \eta_{13}^W= -12\alpha\ ,
\end{array}
\en
with
\be
\alpha=1/(384\pi^2 F^2)\ .
\en
The above results for $\eta_i^W$ were obtained by using, in the ordinary
sector at $p^4$, the chiral Lagrangian term proportional to $l_4$ which
differs from the form originally used in ref.~\cite{gl84}
\be\lbl{L4orig}
{\cal L}_{l_4}^{({\rm orig})}= {il_4\over 4}\braque{ u^\mu \chi_{\mu-} }
\en
by a term proportional to the equation of motion
\be
{\cal L}_{l_4}= {\cal L}_{l_4}^{({\rm orig})}
+{il_4\over4}\braque{ \hat{\chi}_-(\nabla_\mu u^\mu -{i\over2}\hat{\chi}_-)}\ .
\en
If one uses
${\cal L}_{l_4}^{({\rm orig})}$ then, in the odd-intrinsic-parity sector, the
coefficients with labels 6,7 and 8 are modified
to $\tilde{c}_i^W$~\cite{karolfirst,anantmou}.
The relations between $\tilde{c}_i^W$ and $\cw{i}$
are easily worked out by performing a field redefinition,
\bea\lbl{cwtilde}
&& \tilde{c}_6^W = c_6^W -{N_c\over128\pi^2}{l_4\over F^2}
\nonumber\\
&& \tilde{c}_7^W = c_7^W +{N_c\over256\pi^2}{l_4\over F^2}
\nonumber\\
&& \tilde{c}_8^W = c_8^W -{N_c\over512\pi^2}{l_4\over F^2}\ .
\ena
In the present work, we use the original
${\cal L}_{l_4}^{({\rm orig})}$ in our calculations
but we will express the final result in terms
of $\cwr{i}$ rather than $\tilde{c}_i^{Wr}$,
making use of the relations~\rf{cwtilde}
(which will prove slightly more convenient below when we perform
a matching with the $SU(3)$ expansion).

Returning to the $\pi^0$ decay amplitude,
the contributions from the one-loop Feynman
diagrams can be shown to be absorbed into the re-expression of $F$ into
$F_\pi$~\cite{bijnenscornet,donoghuelin},
the physical pion decay constant at order $p^4$,
such that the decay amplitude including the NLO
corrections reads
\bea\lbl{Anlo}
&& T_{LO+NLO}=  {1\over F_\pi}\Bigg\{ {1\over 4\pi^2}
+{16\over3}{\mpid }\left(-4\cwr{3}-4\cwr{7}+\cwr{11}\right)
\nonumber\\
&&\phantom{T_{NLO}=e^2\Big\{ 4\pi}
+{64\over9}{B(m_d-m_u)}(5\cwr{3} + \cwr{7}+2\cwr{8}) \Bigg\}
\  ,
\ena
where $B=-\lim_{m_u=m_d=0}\braque{\bar{u}u}/F^2$ and
$\mpid$ denotes the mass squared of the neutral pion which,
at this order, is equal to $M^2=B(m_d+m_u)$ .
Eq.~\rf{Anlo} shows that the decay amplitude
at NLO receives  a contribution proportional to the isospin breaking
mass difference $m_d-m_u$.
As can be seen from eqs.~\rf{etadiv}
the two combinations of chiral couplings
which enter into the expression of $T_{NLO}$ are finite.
The expression of $T_{LO+NLO}$ therefore involves no chiral logarithm.
The chiral corrections
to the current algebra result are purely polynomial in $m_u,\ m_d$ and
are controlled by four coupling constants from eq.~\rf{lagwz6}.
In order to estimate quantitatively the effects of the NLO corrections,
we will show below that it is useful to express these couplings
as an expansion in powers of the strange quark mass.
Before doing so, let us now investigate the presence of chiral
logarithms, which could possibly be numerically important, in the NNLO
corrections.

\section{$\pi^0$ decay to  NNLO in the two-flavour expansion}\lblsec{NNLOcalc}
We must calculate now a) the one-loop Feynman diagrams with one vertex
involving an NLO chiral coupling, either $l_i$ or $c_i^W$ and b) the two-loop
Feynman diagrams with one vertex taken from the LO Wess-Zumino
Lagrangian and the other one taken from the $O(p^2)$ chiral Lagrangian.
It is convenient the use the following representation for the chiral field
\be\lbl{U}
U=\sigma+i{\vec{\tau}\cdot\vec{\pi}\over F},\quad
\sigma=\sqrt{1-{\vec{\pi}^2\over F^2}}
\en
(since, in this representation, there is no $\gamma4\pi$ vertex at LO).
At the order considered, all
the reducible diagrams are generated from wave-function
renormalization. The expression for the WF renormalization constant $Z$
(corresponding to~\rf{U}) was first given by B\"urgi~\cite{burgi},
\bea\lbl{z12exp}
&& Z^{1\over2}= 1 -{T_M\over 2 F^2} +{1\over F^4} \Bigg[ -{1\over8}T_M^2
+{M^4\over2}
\left(r_Z + \dot{T}_M^2 Q^Z -\dot{T}_M \sum_{i=1}^3 l_i Q_i^Z\right)
\nonumber\\
&&\phantom{Z^{1\over2}= 1 -{T_M\over 2 F^2} }
+{B^2(m_d-m_u)^2}\left( -8F^2(c_7+c_9)+\dot{T}_M l_7\right)
\Bigg]
\ena
with
\be
T_M= {(M^2)^{{d\over2}-1} \Gamma(1-{d\over2})\over (4\pi)^{d\over2} },\
\dot{T}_M=  {dT_M\over dM^2},\
d=4+2w\ .
\en
We have indicated explicitly here the contributions proportional to
$(m_d-m_u)^2$ for completeness because isospin breaking contributions
play an important role for the $\pi^0$ decay amplitude.
We will also need the expression for the chiral expansion of $F_\pi$
at order $p^6$ (from~\cite{bijetalpipi2})
\be\lbl{fpiexp}
{F_\pi\over F}= 1 +{1\over F^2} [ M^2 l_4 -T_M ]
+{M^4\over F^4}
\left[r_F + \dot{T}_M^2 Q^F -\dot{T}_M \sum_{i=1}^4 l_i Q_i^F\right]
+{8B^2(m_d-m_u)^2\over F^2}(c_7+c_9) \ .
\en
The numerical parameters $Q^Z$ and $Q^F$ which appear above read
\be
Q^Z= {1\over96}(96-464 w +1185 w^2),\quad
Q^F=-{1\over192}(240-656 w +1125 w^2)
\en
and will need the following relations obeyed by the numerical
parameters $Q_i^Z$ and $Q_i^F$
\be
Q_1^F= {-1\over2} Q_1^Z\quad  Q_2^F= {-1\over2} Q_2^Z,\quad
Q_3^F= Q_3^Z=2,\quad Q_4^F ={1\over2(1+w)}\ .
\en
Finally, the entries $r_Z$ and $r_F$ in eqs.~\rf{z12exp},~\rf{fpiexp}
represent combinations of coupling
constants from the $O(p^6)$ chiral Lagrangian. The $\pi^0$ amplitude
involves the combination $r_Z+2 r_F$ which is expressed in terms of a
single $p^6$ coupling, called $c_6$ in the classification of
ref.~\cite{bce99a}
\begin{figure}[t]
\centering
\begin{tabular}{cccc}
\includegraphics[width=3cm]{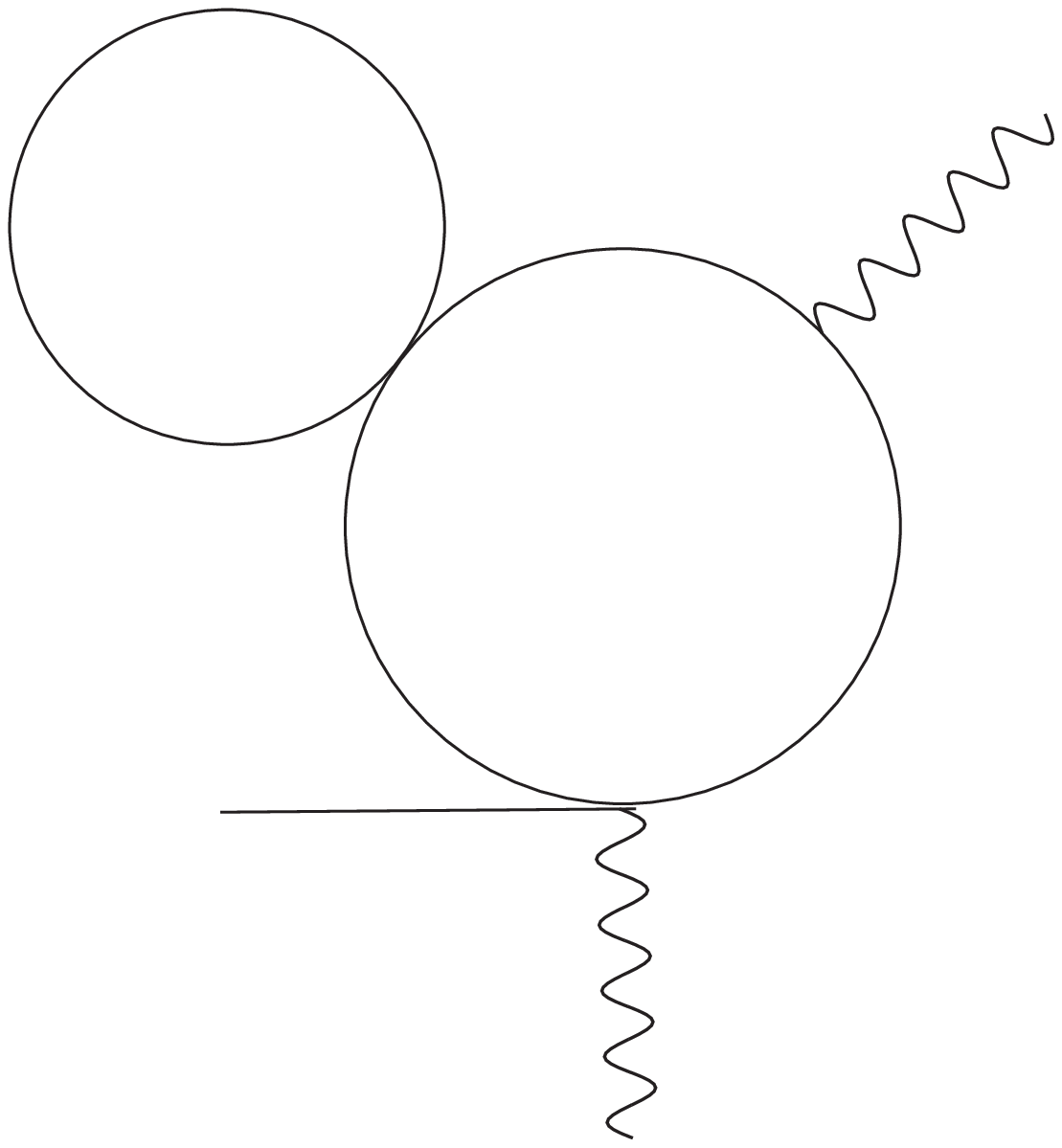} &
\includegraphics[width=3cm]{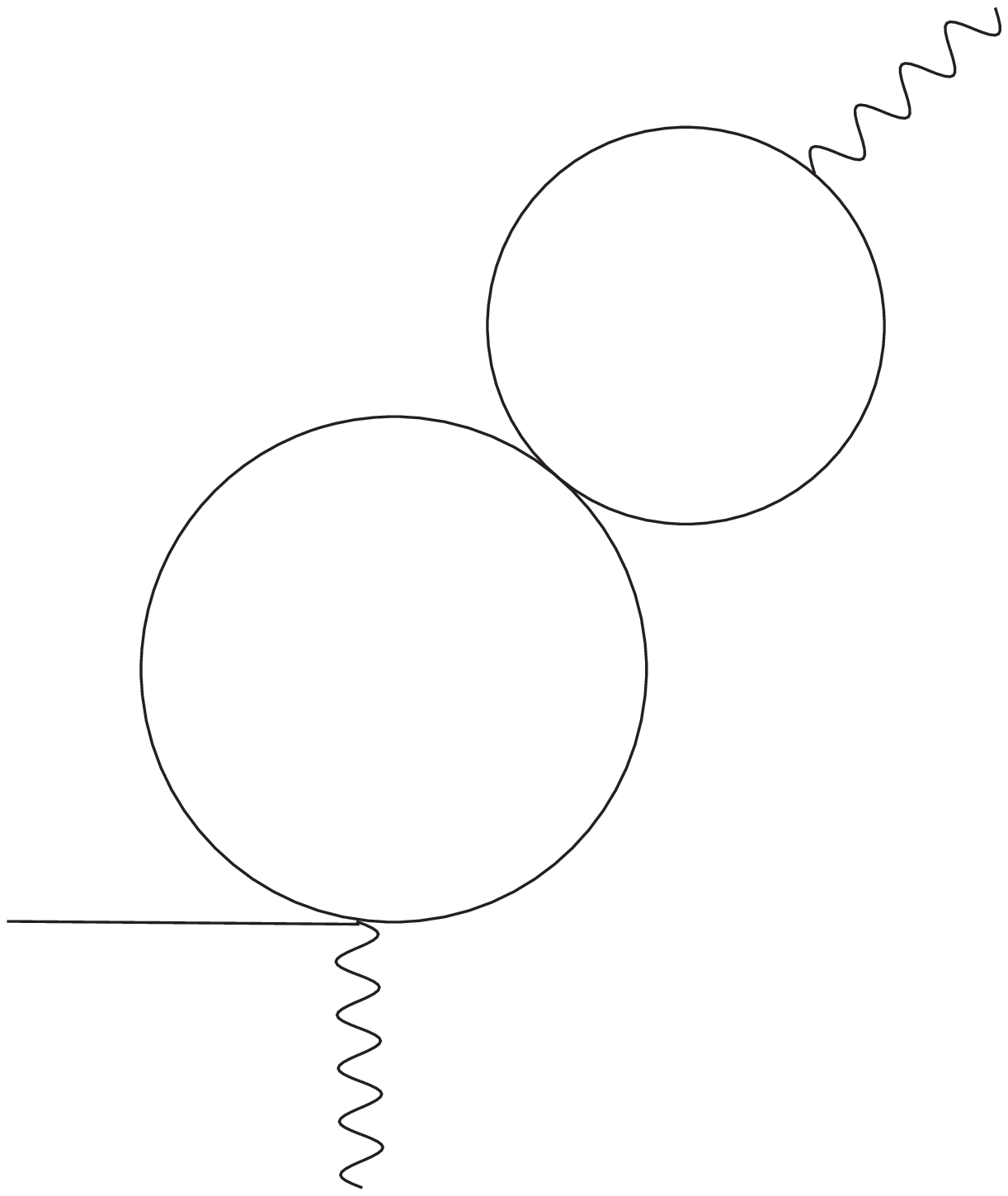} &
\includegraphics[width=1.8cm]{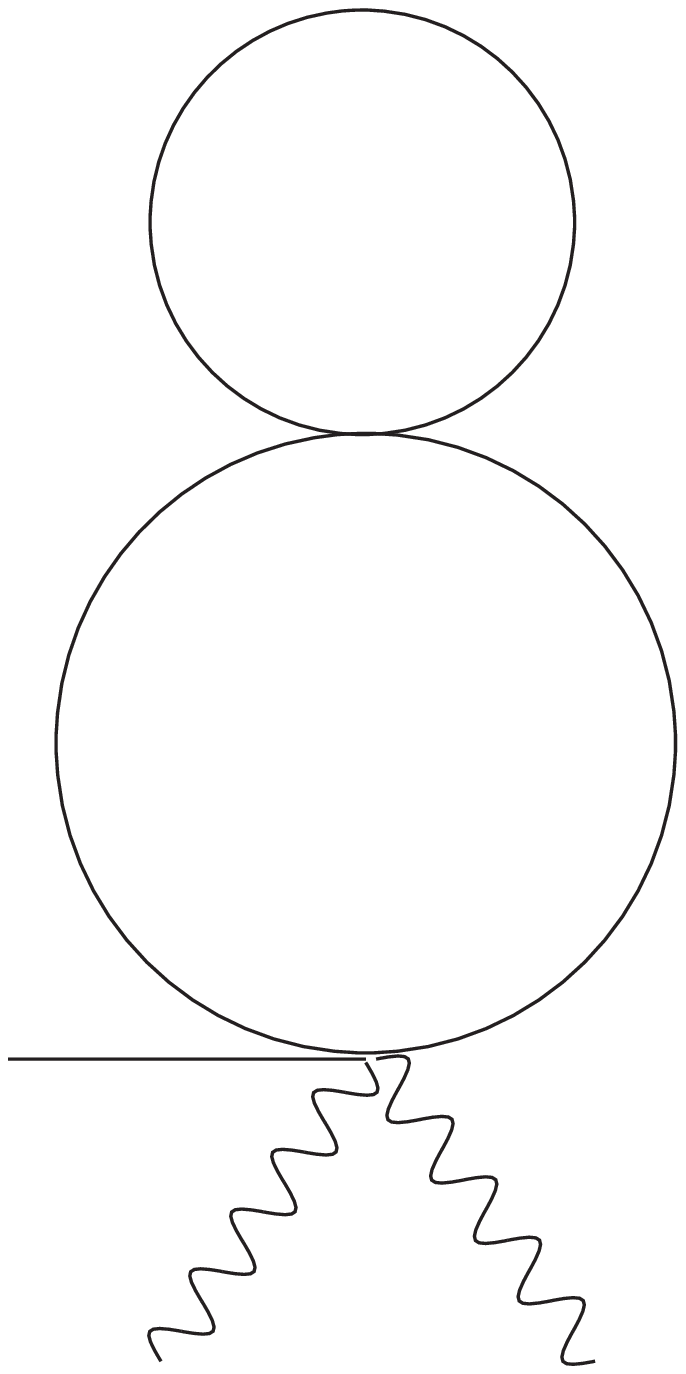} &
\includegraphics[width=2cm]{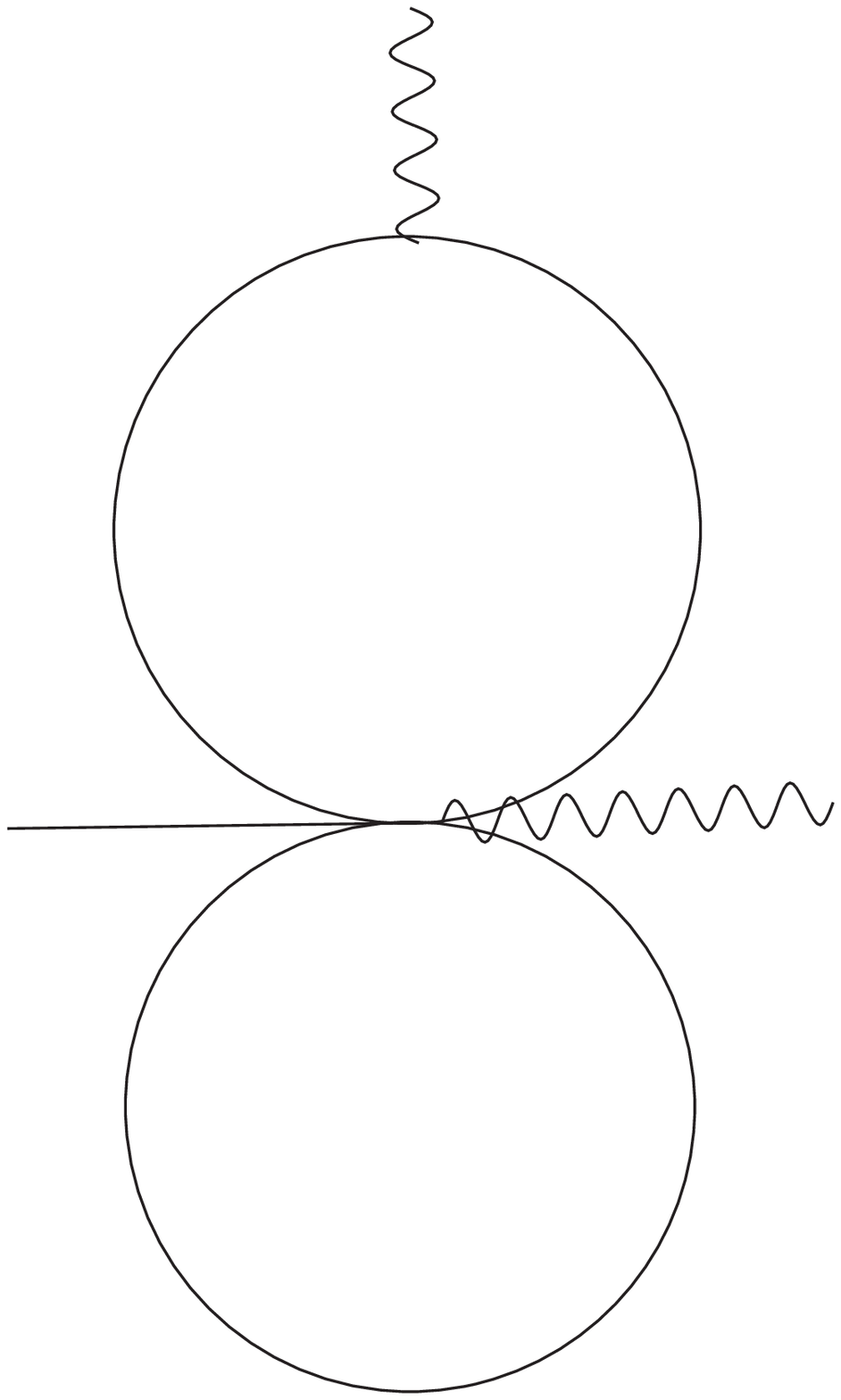} \\
\Large{\bf (a)} & \Large{\bf (b)} & \Large{\bf (c)} & \Large{\bf (d)}\\[0.3cm]
\includegraphics[width=2.5cm]{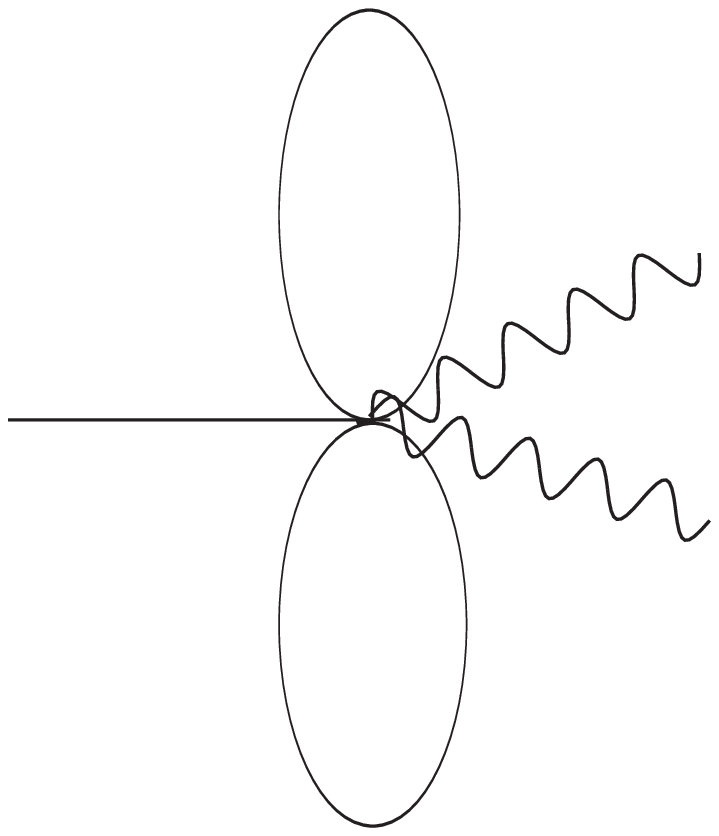} &
\includegraphics[width=3.5cm]{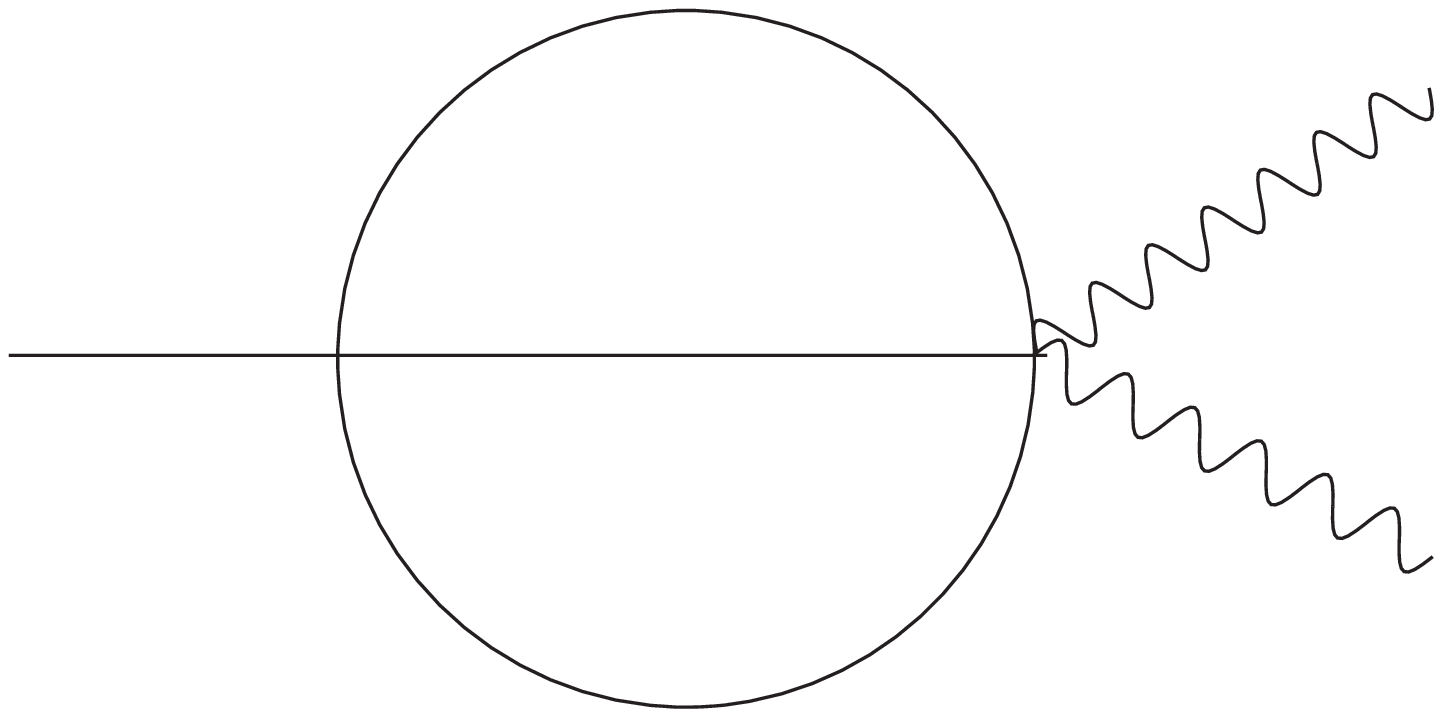} &
\includegraphics[width=3.5cm]{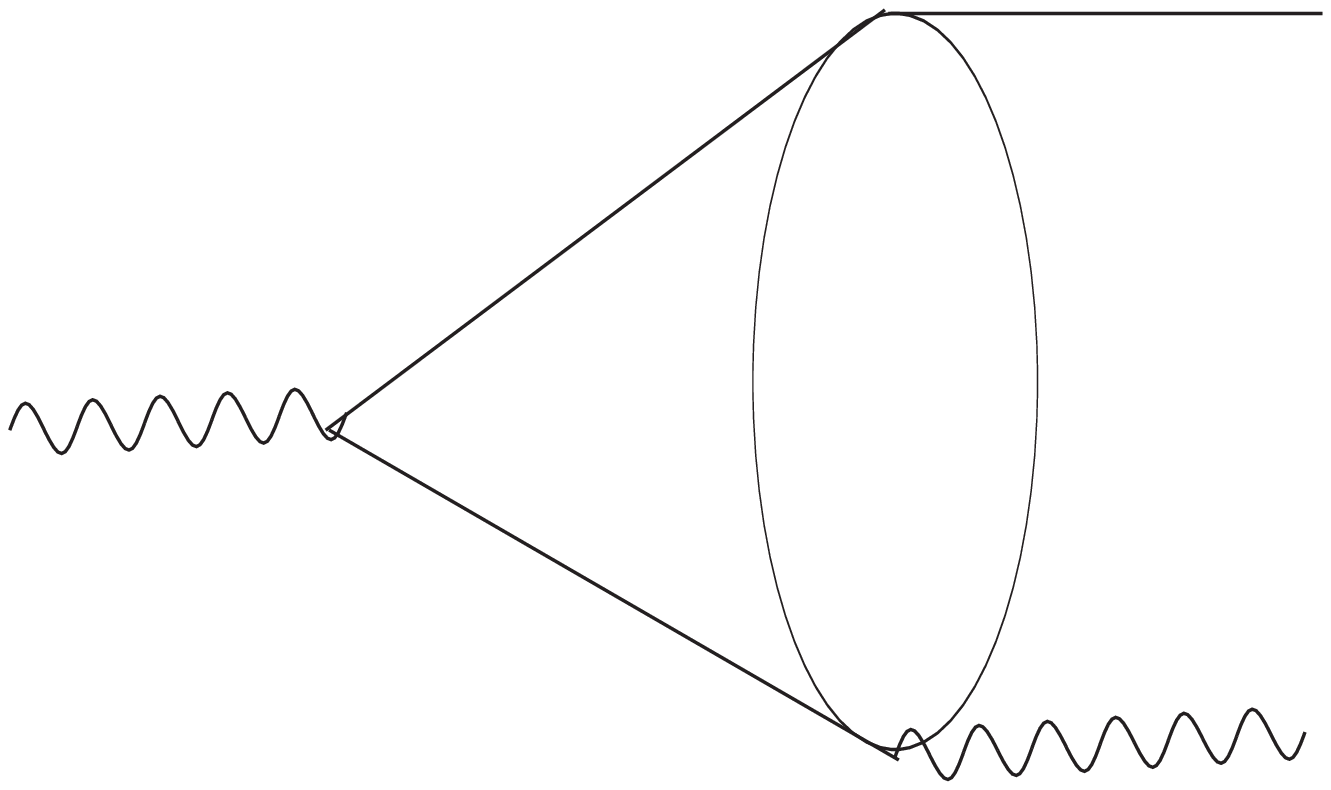} \\
\  \\
\Large{\bf (e)} & \Large{\bf (f)} & \Large{\bf (g)} & \     \\
\end{tabular}
\caption{\sl Two-loop Feynman graphs (one-particule irreducible)
contributions to the $\pi^0\to2\gamma$ amplitude.}
\lblfig{2loopsdiags}
\end{figure}
\be
r_Z+2 r_F = -64 F^2\,c_6\ .
\en
The two-loop one-particle irreducible diagrams which one must compute
(using the representation~\rf{U}) are shown in fig.~\fig{2loopsdiags}.
It turns out to be possible to express all of them analytically in terms
of known special functions by combining the methods exposed in
ref.~\cite{gassersainio} with integration by parts methods. We give the results
corresponding to the two diagrams (f) and (g), which are the most
difficult ones, in appendix 1.

Collecting all the pieces together, we find that the expression for the
NNLO contribution to the $\pi^0$ decay amplitude into two photons has the
following expression,
\bea\lbl{finalNNLO}
&& F_\pi \,T_{NNLO}=
-{M^4\over24\pi^2 F^4}\,\left( {1\over16\pi^2}L_\pi \right)^2
\nonumber\\
&& + {M^4\over 16\pi^2 F^4}\,L_\pi
\left[ {3\over256\pi^4} +
{32F^2\over3}\left(2\cwr{2}+4\cwr{3}+2\cwr{6}+4\cwr{7}-\cwr{11}\right)
\right]
\nonumber\\
&&
+{32 M^2 B(m_d-m_u)\over 48\pi^2 F^4}\,L_\pi\,
\left[-6\cwr{2}-11\cwr{3}+6\cwr{4}-12\cwr{5}-\cwr{7}-2\cwr{8}\right]
\nonumber\\
&& +{M^4\over F^4} \lambda_+   +{M^2 B(m_d-m_u) \over F^4} \lambda_-
+{B^2(m_d-m_u)^2\over F^4} \lambda_{--}\ ,
\ena
where $L_\pi$ represents the chiral logarithm
\be
L_\pi=\log{\mpid\over\mu^2}
\en
and $\lambda_+$, $\lambda_-$, $\lambda_{--}$
can be expressed as follows in terms of renormalized
chiral coupling constants,
\bea
&& \lambda_+ = {1\over\pi^2}\left[
-{2\over 3} d_+^{Wr}(\mu) -8c_6^r  -{1\over4}(l_4^r)^2
+{1\over 512\pi^4} \left( -{983\over288} - {4\over3}\zeta(3)
+3\sqrt3\, {\rm Cl}_2(\pi/3)
\right)\right]
\nonumber\\
&& \phantom{\lambda_+ = } +{16\over3} F^2
\left[\,8l_3^r(\cwr{3}+\cwr{7})+l_4^r(-4\cwr{3}-4\cwr{7}+\cwr{11})\right]
\nonumber\\
&& \lambda_-= {64\over9}\left[
d_-^{Wr}(\mu) +F^2 l_4^r\,(5\cwr{3}+\cwr{7}+2\cwr{8}) \right]
\nonumber\\
&& \lambda_{--}= d_{--}^{Wr}(\mu)-128F^2 l_7 (\cwr{3}+\cwr{7})\ .
\ena
Here, the notation $d^{Wr}$ refer to combinations of couplings from
the NNLO Lagrangian (i.e. of order $p^8$) in the anomalous sector.

A few remarks are in order concerning this calculation.
First, concerning non-local
divergences, i.e. terms of the form $M^4\log(M^2)/(d-4)$, we have
verified that those which are generated from the two-loop diagrams are
cancelled exactly by those generated from the one-loop diagrams proportional to
$l_i$, $c_i^W$ as expected from the Weinberg consistency conditions. The
divergences that are left are proportional to $M^4$, $M^2(m_d-m_u)$
and $(m_d-m_u)^2$. They are cancelled by
the contributions, at tree level, from the chiral Lagrangian of order
$p^8$ in the anomalous sector.
We have denoted the three independent combinations
of $O(p^8)$ chiral couplings by $d_+^W$, $d_-^W$ and $d_{--}^W$.
Our calculation shows
that the relation between  these and the corresponding renormalized
combinations must be as follows,
\bea
&& d_+^W = {(c\mu)^{2(d-4)}\over F^2} \Bigg[
d_+^{Wr}(\mu) -\Lambda^2\left(-{17\over3}\right)
-\Lambda\bigg(
-11 l_1^r-7 l_2^r-{1\over2}l_3^r-{3\over2}l_4^r-{53\over4608 \pi^2}
\nonumber\\
&&\phantom{ d_+^{Wr} -\Lambda^2\left(-{17\over3}\right)}
+16\pi^2 F^2\left(-4\cwr{2}-2\cwr{3}-4\cwr{6}-2\cwr{7}+\cwr{11} \right)
\bigg) \Bigg]
\nonumber\\\
&& d_-^W = {(c\mu)^{2(d-4)}\over F^2}\left[d_-^{Wr}(\mu)
-\Lambda F^2 \left(-18\cwr{2}-23\cwr{3}+18\cwr{4}-36\cwr{5}-\cwr{7}-2\cwr{8}\right)
\right]\nonumber\\
&& d_{--}^W = {(c\mu)^{2(d-4)}\over F^2}\left(  d_{--}^{Wr}(\mu)
+{\Lambda l_7\over\pi^2} \right)\ .
\ena
Eq.~\rf{finalNNLO} shows that chiral logarithms are indeed present at NNLO. The
coefficient of the dominant one, as can be shown quite generally,
depends only on $F$. The coefficient of the subdominant chiral logarithm
has one part depending only on $F$ and another one depending on the NLO chiral
couplings $\cwr{i}$. From a numerical point of view, the contribution from
the dominant chiral logarithm turns out to be very small, of the order of a few
per mille. This lack of enhancement could indicate a fast convergence
of the chiral perturbation series. In this respect, the detailed
formula~\rf{finalNNLO} could be used in association with results from lattice
QCD simulations, in which the quark masses $m_u$, $m_d$ are larger than the
physical ones and can be varied.
This would allow one to determine the relevant combinations of chiral
couplings. In the following section we discuss an alternative,
more approximate method, to estimate these combinations.

\section{Chiral expansion in $m_s$}
From now on, we assume that the mass of the strange quark is sufficiently
small, such that the chiral expansion in $m_s$ is meaningful.
One can then calculate the $\pi^0$ lifetime using the
three-flavour  chiral expansion.
Instead of doing so directly, as it remains true that $m_u,\ m_d <<
m_s$,  it is instructive to start from the $SU(2)$ expression,
eq.~\rf{finalNNLO}  and perform a chiral expansion of the couplings $c_i^W$
as a function of $m_s$. A priory, one expects expressions of the
following form to arise
\be\lbl{ciwmsexp}
\cwr{i} = {\alpha_i\over m_s} +
\left(
\beta_i + \sum_a \gamma_{ia}\CWr{a} +\delta_i \log{ B_0 m_s\over \mu^2}
\right)+ O(m_s)\ ,
\en
where $\CWr{a}$, $a=1\dots24$ are the coupling constants of the NLO Lagrangian
in the anomalous sector in the $SU(3)$ expansion~\cite{girlanda}
and $B_0=\lim_{m_s=0} B$.
Analogous expansions were established in ref.~\cite{gl85} for the
$SU(2)$ couplings $B$, $F$ and $l_i^r$. This problem was reconsidered recently
in ref.~\cite{gasserhaefeli} in which the NNLO terms
in that expansion have been derived.
Also in ref.~\cite{haefeli} the $m_s$
expansions of the $SU(2)$ LEC's in the electromagnetic sector were studied.
In order to generate such expansions one can work in the
$SU(2)$ chiral limit $m_u=m_d=0$,
compute sets of correlations functions
having $SU(2)$ flavour structure in both the $SU(2)$ and the $SU(3)$ chiral
expansions and equate the expressions. The authors of ref.~\cite{gasserhaefeli}
have shown how to perform this matching at the level of the generating
functionals.  In the $SU(3)$ generating functional, one must
use external sources $s$, $p$, $v_\mu$, $a_\mu$ which correspond to those
used in the $SU(2)$ functional embedded into $3\times 3$ matrices.
Since there is no source for strangeness,
the classical $SU(3)$ chiral field involves the three pions $\pi_a$ and
the $\eta$ field but no kaons
\be\lbl{U_cl}
U_{cl}= \exp{ i\lambda_a\pi_a\over F_0}\exp{i\eta\lambda_8\over F_0}
\en
($F_0$ being the pion decay constant in the three-flavour chiral limit).
Using the equation of motion one can express the field $\eta_{cl}$ in terms
of an $SU(2)$ chiral building-block~\cite{gasserhaefeli,haefeli}
\be\lbl{etaclass}
{\eta_{cl}\over\sqrt3 F_0}= i\braque{\chim}\left(-{1\over 16 m_s B}\right)
+O(p^4)\ .
\en
The terms proportional to $\eta_{cl}$ thus generate contributions proportional
to $1/(m_s B)$. These can be also seen as resulting from $\eta$ meson
propagators in tree diagrams. Besides, eq.~\rf{etaclass} shows that
$\eta_{cl}$ counts as $O(p^2)$ in the $SU(2)$ counting. Inserting
$U_{cl}$ from eq.~\rf{U_cl} in the $SU(3)$ Wess-Zumino action and
expanding to first order in $\eta_{cl}$ we obtain,
\bea\lbl{etaWZ}
&& {\cal L}_\eta= -{iN_c\over 48\pi^2} {\eta_{cl}\over \sqrt3 F_0}
\epsilon^{\mu\nu\alpha\beta}\,\bigg\{
 {1\over2} \braque{\fpab\umu\unu}
-{3\over8}i\braque{\fpab\fpmn}
\nonumber\\
&&\phantom{ {\cal L}_\eta= -{iN_c\over 48\pi^2} }
+{3\over4}i\braque{\fpab}\braque{\fpmn}
-{1\over8}i\braque{\fmab\fmmn} \bigg\}\ .
\ena
This allows one to deduce the leading terms, which behave as $1/m_s$,
in the expansion of
the couplings $\cwr{i}$. Next,
the terms proportional to $(m_s)^0$ are generated from
three sources.
\begin{itemize}
\item[1)] From the $SU(3)$ Lagrangian ${\cal L}_6^{W}$,
by inserting $U_{cl}$ (with $\eta_{cl}$ set to zero), which gives contributions
proportional to LEC's $\CWr{i}$.
\item[2)] From one-loop irreducible graphs
with one vertex taken from the Wess-Zumino action
and having one kaon or one eta running in the loop.
\item[3)] From corrections to the $\eta$
pole contributions stemming from tadpoles or from vertices proportional
to the $O(p^4)$ couplings $L_i$.
\end{itemize}
The results are presented in
eqs.~\rf{ciwexp1} below and ~\rf{ciwexp3} in the appendix.

Let us now examine the applications of this exercise to the problem
of the $\pi^0$ lifetime.
As seen in sec.~\sect{SU2exp} the NLO corrections involve two independent
pieces, one proportional to $\mpid$ and one to $B(m_d-m_u)$, and they are
controlled by two combinations of the four couplings
$\cwr{3}$,  $\cwr{7}$, $\cwr{8}$ and $\cwr{11}$.
For these, we take into account the first two terms in the $m_s$ expansion
which read
\bea\lbl{ciwexp1}
&&\cwr{3} = -{3\over2}c_0+\CWr{7} + 3 \CWr{8} +O(m_s)
\nonumber\\
&& \cwr{7} =\phantom{-}{3\over2}c_0 -3\CWr{8} + {1\over4} \CWr{22} +O(m_s)
\nonumber\\
&& \cwr{8} =\phantom{-}{3\over4}c_0 + {1\over2} \CWr{7}+ 3 \CWr{8}
- {1\over8} \CWr{22} +O(m_s)
\nonumber\\
&& \cwr{11} =\phantom{-2c_0+} \CWr{22} +O(m_s)\ ,
\ena
where
\be\lbl{c0}
c_0= {1\over32\pi^2}\left[ -{1\over16 B m_s}+
{2\over F_0^2}\left( 3\Lr{7}+\Lr{8}
-{1\over512\pi^2}(L_K+{2\over3}L_\eta) \right) \right]
\en
and
\be\lbl{Lk}
L_K=\log{m_s B_0\over\mu^2},\quad L_\eta=L_K+\log{4\over3}\ .
\en
At this point, one observes that by using the $m_s$ expansion, we
have expressed four $SU(2)$ couplings in terms of three $SU(3)$ ones.
This might look as a modest improvement. Fortunately, the combinations
relevant for the $\pi^0$ lifetime at NLO actually involve only two
couplings $\CWr{7}$, $\CWr{8}$ while $\CWr{22}$ drops out.

Let us now consider the terms proportional to $m_\pi^4$ and $\mpid(m_d-m_u)$.
One can see from eq.~\rf{finalNNLO} that they involve four more LEC's,
$\cwr{2}$, $\cwr{4}$, $\cwr{5}$, $\cwr{6}$. It makes sense here to retain
only the part of these LEC's which are dominant in the $m_s$ expansion,
i.e. the $1/m_s$ part,
\be\lbl{ciwexp2}
 \cwr{2}\simeq  \tilde{c}_0,
 \quad
 \cwr{4}\simeq -{1\over2}\tilde{c}_0,
 \quad
 \cwr{5}\simeq  0,
 \quad
 \cwr{6}\simeq -\tilde{c}_0,\quad
\tilde{c}_0=-{1\over 512 \pi^2 B m_s}
\en
and we perform a similar approximation in eq.~\rf{ciwexp1}.
We will also retain the part involving the LEC $\CW{8}$ as it will appear
that the size of this coupling is comparable to that of the $1/m_s$ terms.
Inserting the $m_s$ expansions ~\rf{ciwexp1} and ~\rf{ciwexp2},
in the $SU(2)$  chiral expansion of the $\pi^0$ decay amplitude~\rf{finalNNLO}
we obtain the following expression
\bea\lbl{Api0NLO+}
&& T_{(LO+NLO)_+}= {1\over F_\pi}\Bigg\{
{1\over 4\pi^2}
-{64\over3}\mpid\CWr{7}
+{1\over 16\pi^2} {m_d-m_u\over m_s}
\Big[ 1-{3\over2}{\mpid\over16\pi^2\fpid}L_\pi \Big]
\nonumber\\
&&\phantom{T_{NLO+}}
+32B(m_d-m_u)\Bigg[  {4\over3}\CWr{7}
+ 4\CWr{8}\Big(1 -3  {\mpid\over16\pi^2\fpid}L_\pi \Big)
\\
&& \phantom{T_{NLO+}}
-{1\over16\pi^2 F_\pi^2} \Big( 3L_7^r +L_8^r
-{1\over512\pi^2}(L_K+{2\over3}L_\eta) \Big)  \Bigg]
-{1\over24\pi^2} \left( {\mpid \over16\pi^2 \fpid}L_\pi\right)^2
\Bigg\}\nonumber\ .
\ena
\subsection{A modified $SU(3)$ chiral counting}\lblsec{modifSU3}
Some comments are in order concerning eq.~\rf{Api0NLO+}.
In particular, one expects it to be related to the formula
that one can compute starting from $SU(3)$ ChPT.
Such a computation was performed, e.g. in ref.~\cite{anantmou}.
In $SU(3)$ ChPT $m_u,\ m_d$ and $m_s$ are counted on the same footing,
\be
m_u,\ m_d \sim m_s \sim O(p^2)\qquad [\,{\rm standard}\ SU(3)]\ .
\en
In the physical situation, however, $m_u,\ m_d << m_s$. For processes which
involve only pions this can be accounted for by adopting the following
modified counting,
\be\lbl{modSU3}
m_u,\ m_d \sim O(p^2),\quad m_s\sim O(p)\qquad  [\,{\rm modified}\ SU(3)]\ .
\en
The formula~\rf{Api0NLO+} for the $\pi^0$ lifetime can be argued to be
a consistent expansion in this modified counting.
One notes first that all the corrections must be proportional to
$m_u, m_d$ since the starting point is exact in the SU(2)
chiral limit. The formula~\rf{Api0NLO+} includes the leading corrections
of order $p$ (which must be proportional to $m_u/m_s$, $m_d/m_s$) as well as
the subleading corrections of order $p^2$ (which must be
proportional to $m_u$, $m_d$).
It also includes the corrections of order
$p^3$ which are logarithmically enhanced (which must be proportional
to $m_u m_s$, $m_d m_s$ multiplied by $\log(m_u+m_d)$
as well as the corrections of order $p^4$ which are double logarithmically
enhanced.
Obviously, by retaining logarithmically enhanced terms at a given order
instead of the full set of terms, one introduces a chiral scale dependence
into the amplitude. Clearly, one should use a value of the scale of
the order of the kaon or the eta mass for this approximation to make sense.
Finally,
we have verified that, starting from the expression for the amplitude in
standard $SU(3)$ at NLO obtained in ref.~\cite{anantmou},
and expanding in powers of $m_u/m_s$, $m_d/m_s$ one recovers
exactly the terms of order
$p$, $p^2$ and $p^3\log(p^2)$ in the modified $SU(3)$ expansion~\rf{modSU3}.
In practice, the expression~\rf{Api0NLO+} is somewhat simpler than the
standard $SU(3)$ NLO expression and contains the double logarithm term.
The latter turns out to be numerically small so that the two expressions
are essentially equivalent in practice. In order to derive a numerical
prediction from eq.~\rf{Api0NLO+} one needs inputs for:
$F_\pi$, $(m_d-m_u)/m_s$, $B(m_d-m_u)$ and $\CW{7}$, $\CW{8}$.
We will give an update on the determination of these quantities
in sec.~\sect{updates}

In addition to the chiral corrections induced by the quark masses,
one should also take electromagnetic corrections into account. These have
been considered in ref.~\cite{anantmou},
where the correction terms of order $e^2$
and of order $e^2 (m_u+m_d)/m_s$ have been computed. Here, it is consistent
to retain only the term of order $e^2$, its expression in terms
of Urech's chiral couplings~\cite{urech95} is recalled,
\be
T_{e^2}= {e^2\over4\pi^2 F_\pi}
\left\{ -{4\over3}(K_1^r+K_2^r)+2K_3^r-K_4^r-{10\over9} (K_5^r+K_6^r)
+{C\over 32\pi^2 F_\pi^4}(5 + 4L_\pi +L_K)
\right\}\ .
\en
This term is defined such that the $\pi^0\to2\gamma$ amplitude
is expressed in terms of $F_{\pi^0}$ which is the neutral pion
decay constant in {\it pure QCD}
and $m_\pi^2$ which is the {\it physical} neutral pion mass
(i.e. including EM corrections).
\section{Phenomenological updates}\lblsec{updates}
Let us now update the various inputs
needed to calculate the numerical prediction for the $\pi^0$
lifetime in ChPT.
\begin{itemize}
\item[1)] $F_\pi$:\\
An obviously essential input here is $F_\pi$, the value of the pion decay
constant.
Marciano and Sirlin~\cite{marcianosir}
have evaluated the radiative corrections in the process
$\pi^+\to \mu^+\nu (\gamma)$ decay rate
such that it is expressed in terms of
$F_{\pi^+}$ the charged pion decay constant in pure QCD.
In pure QCD the difference between $F_{\pi^0}$ and $F_{\pi^+}$ is quadratic
in the quark mass difference $m_d-m_u$ and can be expressed as follows
in ChPT,
\be\lbl{Fpiisob}
\left.{F_{\pi^+}\over F_{\pi^0}}\right\vert_{QCD}-1
= {B^2(m_d-m_u)^2\over F_\pi^4}\left[
-16\,c_9^r(\mu) -{l_7\over 16\pi^2}\left( 1+\log{\mpid\over\mud}\right)
\right]\simeq 0.7\,10^{-4}\ .
\en
A rough numerical evaluation has been made
by using leading order $1/m_s$ estimates
\be
l_7\simeq{\fpid\over 8Bm_s},\quad c_9^r\simeq -{3\over2}\left(
{\fpid\over Bm_s }\right)^2\ .
\en
Eq.~\rf{Fpiisob} shows that the difference between $F_{\pi^+}$
and $F_{\pi^0}$ is negligibly small for our purposes, and we will ignore it.
In the expression of ref.~\cite{marcianosir} for the radiative corrections,
one constant term, called  $C_1$, was left undetermined.
Matching with the ChPT expansion of
the $\pi^+$ decay rate at $O(p^4)$ one can express $C_1$ in terms of chiral
logarithms and a set of chiral couplings~\cite{neufeldknecht}. The latter
can then be estimated using chiral sum rules and resonance
saturation~\cite{descotesmou}. Using these results and the updated value
of $V_{ud}$ from ref.~\cite{townerhardy}
\be
V_{ud}=0.97418(26)\ ,
\en
we find
\be
F_\pi= 92.22\pm0.07\ {\rm MeV}\ .
\en
\item[2)]{$B(m_d-m_u)$, $(m_d-m_u)/m_s$, $3L_7+L_8^r$}:\\
Because of the Kaplan-Manohar invariance~\cite{kaplanman} it is not possible
to determine independently the quark mass ratios and the couplings
$L_7$, $L_8$ in ChPT using low-energy data. One may use an input
from lattice QCD, e.g. on the quark mass ratio $r= 2m_s/(m_u+m_d)$.
Using the results obtained in ref.~\cite{allton2+1}
as well as those from other recent QCD simulations which are collected
in table XVI of that reference and averaging, one can deduce
\be\lbl{rval}
r\equiv {2m_s\over m_u+m_d}= 28.0\pm1.5\ .
\en
Using this input for $r$, we may treat terms linear in the quark masses in
NLO ChPT expressions as follows,
\be
(m_u+m_d)B_0\simeq \mpid,\quad m_s B_0 = {r\over 2}\mpid\ .
\en
The value of the LEC combination $3L_7+L_8^r$,
can be deduced using $r$ and
standard $O(p^4)$ ChPT formulas for the pseudo-scalar meson
masses~\cite{gl85}
\be
3L_7+L_8^r(\mu)= (0.10\pm0.06)\,10^{-3}\quad (\mu=M_\eta)\ .
\en
Concerning the quark mass difference $m_d-m_u$, we will use the recent
determination made in ref.~\cite{bijnensghorbani}.  It is based on the
$\eta\to 3\pi$ decay amplitude which is an isospin breaking observable with
very small sensitivity to electromagnetic
effects~\cite{kamboreta3piem,kubiseta3piem}.
The amplitude has been computed at order $p^6$ in ChPT  by the authors of
ref.~\cite{bijnensghorbani} and they deduce the following result\footnote{%
An alternative evaluation of $R$ can be made based on the $K^+-K^0$
mass difference. As one can see from table 6 of ref.~\cite{bijnensghorbani}
this method tends to give values of $R$ smaller than eq.~\rf{Rval}. The
calculation of the $K^+-K^0$ mass difference in ChPT, however, has
uncertainties related to the couplings $C_i$ and also from estimates of the
electromagnetic contributions, beyond the Dashen low-energy theorem, which
have some model dependence.
One could also use isospin violation in $K_{l3}$ form factors. For an updated
discussion of these effects see~\cite{Kastner-Neufeld}.
}
,
\be\lbl{Rval}
R\equiv {m_s-\hat{m}\over m_d-m_u}=42.2\
\en
(with $\hat{m}=(m_u+m_d)/2$).
No figure for the uncertainty is given.
We have estimated it by noting that the main source of uncertainty
in this result comes from the unknown values of the coupling constants $C_i^r$
from the $O(p^6)$  Lagrangian. For these couplings, it was shown that
simple resonance models are sometimes misleading~\cite{karoletmoi}
because of their strong scale dependence.
We have estimated the order of magnitude of the uncertainty by taking the
difference between the value of $R$ obtained from a $p^6$ calculation
and the value obtained from a $p^4$ calculation and dividing by two, which
gives
\be\lbl{deltaR}
\Delta R\simeq 5\ .
\en
Using~\rf{rval},~\rf{Rval} and \rf{deltaR}, we obtain\footnote{
In ref.~\cite{deandrea} a determination of the quantity $B_0(m_d-m_u)$
from $\eta\to3\pi^0$ was proposed, based on using both the decay rate and the
slope parameter $\alpha$, obtaining
$B_0(m_d-m_u)=(0.25\pm0.02)\,M^2_{\pi^0}$. This appears somewhat
smaller than the result in eq.~\rf{md-mu} but one should keep in mind
that the ratio $B_0/B$, while expected to be close to one, is not
accurately known.
}
\be\lbl{md-mu}
{m_d-m_u\over m_s}= (2.29\pm0.23)\,10^{-2},\qquad
B(m_d-m_u)= (0.32\pm 0.03)\,M^2_{\pi^0}\ .
\en
\item[3)]$\CW{7}$:\\
This constant obeys a sum rule in terms of the form factor associated
with the photon-photon matrix element of the pseudoscalar
current~\cite{kitazawa}. A simple resonance saturation approximation
in this sum rule gives a relation between $\CW{7}$ and
the $\pi(1300)$ mass and its couplings to the
pseudoscalar current ($d_m$) and to
two photons ($g_{\piprim}$)~\cite{anantmou}
\be
\CW{7}\simeq {g_{\piprim} d_m\over M^2_{\pi'}}\ .
\en
Recent experimental data by the Belle collaboration has confirmed the
extreme smallness of the coupling of the $\pi(1300)$ meson to two
photons~\cite{bellepiprime}
\be
\Gamma_{\pi'\to 2\gamma} < 72\ {\rm eV}\ .
\en
The validity of the resonance saturation approximation in this case
might be questioned since, in the sum rule,
$\CW{7}$, could pick up more important contributions
from energies higher than the mass of the $\pi(1300)$ resonance.
There has been several attempts at estimating this high energy
contribution to $\CW{7}$ in the literature:
Using a quark-hadron duality picture,
Kitazawa~\cite{kitazawa} argue that this contribution arises from a
triangle diagram and should thus be proportional to the constituent quark
mass (this result was applied to $\eta$ decay in ref.~\cite{pham}). In
QCD, one expects the constituent quark mass to be momentum dependent
(see e.g.~\cite{terning}) and to decrease at high momenta,
which is not taken into account in this evaluation.
A calculation of the triangle diagram in the NJL model was performed
in ref.~\cite{bijprad94}.  As this model implements a momentum cutoff,
however, it rather concerns the low-energy rather than the high-energy
contribution to $\CW{7}$.
An alternative idea was proposed in ref.~\cite{moussall95} based on a
minimal resonance saturation modelling of the three-point function VVP
and enforcing a correct asymptotic matching to the OPE expansion of
this three-point function. The result, unfortunately, cannot be shown
to remain stable under inclusion of more resonances.
None of the estimates, finally, appear to be quantitatively very
compelling.
It seems however quite safe to assume that the coupling $\CW{7}$ should be
suppressed, say by one order of magnitude,
as compared to the coupling $\CW{8}$.
Indeed, in an analogous sum rule representation, $\CW{8}$ picks up a
strong contribution from the $\eta'$ resonance. We will therefore take
\be\lbl{cw7}
\vert \CW{7}\vert < 0.1\,\vert \CW{8}\vert\ .
\en
\item[4)]$\CW{8}$:\\
Having assumed the validity of $SU(3)$ ChPT, together with the result
~\rf{cw7} of the above discussion  on $\CW{7}$, one can determine $\CW{8}$
from the experimental information on the $\eta\to 2\gamma$ decay
width. According to the PDG\footnote{The PDG now rejects the Primakoff
experiment~\cite{browman} which gave a smaller result. A re-discussion of
that experiment has recently appeared~\cite{rodrigues}.}
~\cite{pdg}
\be\lbl{etawidth}
\Gamma_{\eta\to2\gamma}=0.510\pm  0.026\ {\rm keV}\ ,
\en
while the corresponding amplitude computed in ChPT, including LO and NLO
contributions, reads
\bea\lbl{ampliteta}
&& T_\eta={e^2\over \sqrt3 F_\pi}\bigg[
{F_\pi\over 4\pi^2 F_\eta} (1+x_\eta)
-{64\over 3}\,\mpid \CW{7}
\nonumber\\
&&\phantom{T_\eta={e^2\over \sqrt3 F_\pi}\bigg[}
+{256\over3} (r-1)\mpid\Big( {1\over6}\CW{7} +\CW{8}\Big)
+ O(m_s^2)\bigg]\ ,
\ena
where $x_\eta$ encodes isospin breaking effects
\be
x_\eta= \sqrt3 (-\epsilon_1 + e^2 (\delta_\eta-\delta_1) )\simeq -0.023\ ,
\en
using notations and results from refs.~\cite{gl85} and \cite{anantmou}.
We need an input for $F_\eta$ in eq.~\rf{ampliteta}. Up to
corrections quadratic in $m_s$, $F_\eta$ is linearly related to $F_\pi$
and $F_K$~\cite{gl85},
\be\lbl{feta}
F_\eta= {4 F_K -F_\pi\over 3}
+{\mpid\over96\pi^2 F_\pi}\left[2(r+1)\log{2(2r+1)\over3(r+1)}
-\log{2r+1\over3}        \right] + O(m_s^2)\ .
\en
The review in ref.~\cite{antonelli} quotes the following
result for $F_K$ from averaging over recent experiments on $\pi_{l2}$ and
$K_{l2}$  decays
\be
{F_K\, V_{us}\over F_\pi\, V_{ud}}= 0.27599(59)\ .
\en
Assuming exact CKM unitarity we can deduce $F_K$ and then $F_\eta$
\be
F_K= 109.84\pm 0.63,\quad F_\eta= 118.4 \pm 8.0\qquad ({\rm MeV})\ .
\en
The error on $F_\eta$  is dominated by the $O(m_s^2)$ contributions
in eq.~\rf{feta}. We have estimated that it should be smaller  than the
$O(m_s)$ contribution by a factor of three.
Finally, using these results in conjunction with
eqs.~\rf{etawidth}~\rf{ampliteta} we determine the coupling $\CW{8}$
\be\lbl{cw8num}
\CW{8}=(0.58\pm 0.20)\,10^{-3}\ ({\rm GeV^{-2}})\ .
\en
We have estimated that the uncertainty stemming from unknown $O(m_s^2)$
chiral corrections in the $\eta$ decay amplitude to be of order 30\%
compared to the $O(m_s)$ corrections.
\end{itemize}

\begin{table}[htb]
\centering
\begin{tabular}{|c||ccccc|}\hline
CA & $O(p)$ & $O(p^2)$ & $O(e^2)$ & $O(p^3\log\,p)$ &  $O(p^4\log^2p)$
\T\B \\ \hline
7.76 & 0.09 & 0.29 & -0.05 & 0.005 & -0.004 \\ \hline
\end{tabular}
\caption{\sl Current algebra contribution to the $\pi^0\to 2\gamma$
decay width (in eV) and corrections of various chiral orders
using the modified $SU(3)$ counting.}
\lbltab{contribs}
\end{table}
The numerical results for the current algebra amplitude and the corrections
according to the modified chiral $SU(3)$ counting, using the updated inputs
presented above, are collected in table~\Table{contribs}.
One remarks that the $O(p^2)$ contribution is larger than the $O(p)$ one.
This is induced by the size of the LEC $\CW{8}$. Expressed as a sum rule,
$\CW{8}$ is dominated by the $\eta'$ contribution, which can be
written~\cite{anantmou}
\be\lbl{c8wetap}
\CW{8}\simeq {g_{\eta'}\tilde{d}_m\over \metapdchir}\ ,
\en\lbl{c8wetap1}
where $\metapchir_{\,\eta'} $ is the mass of the $\eta'$ in the chiral limit. In the
large $N_c$ limit one has,
\be
g_{\eta'}={\sqrt6\over128\pi^2 F_0},\
\tilde{d}_m= {F_0\over2\sqrt6},\quad
\CW{8}\simeq {1\over 256\pi^2 \metapdchir\;\;}\ .
\en
The enhancement
of  $\CW{8}$ can then be understood, qualitatively, as a large $N_c$ effect.
In practice, the value
of $\CW{8}$ that one  can estimate using the resonance
saturation formula~\rf{c8wetap} agrees reasonably well with the one deduced
from a ChPT expansion of the $\eta\to2\gamma$ amplitude\footnote{
Our result disagrees with ref.~\cite{ioffe} in which the corresponding
contribution is smaller by one order of magnitude.} (eq.~\rf{cw8num}).
The enhancement of
the $O(p^2)$ contribution is therefore a well understood effect
and does not signal a breakdown of the expansion.
Table~\Table{contribs} shows that
the logarithmically enhanced contributions of order $p^3\log(p)$ and
$p^4\log^2(p)$ are quite small in practice and tend to cancel each other.
Finally, the prediction for the $\pi^0$ decay width reads,
\be\lbl{W2gamma}
\Gamma_{\pi^0\to 2\gamma}= (8.09 \pm 0.11)\ {\rm eV}\ .
\en
The two main sources for the uncertainty are: $m_d-m_u$ ($\pm0.05$) and
$\CW{8}$ ($\pm0.098$). We have added the errors in quadrature. Compared
to ref.~\cite{anantmou} the main modification in the
input is the value of the $\eta\to2\gamma$ width in the PDG.
The branching fraction for the $2\gamma$ decay mode is
$(98.798\pm0.032)\%$~\cite{pdg}
(the most sizable other decay being the Dalitz mode
$\pi^0\to \gamma e^+ e^-$, for review see e.g.~\cite{dalitz}).
Our result, eq.~\rf{W2gamma}, then corresponds to the
following value for the $\pi^0$ lifetime
\be
\tau_{\pi^0}= (8.04\pm 0.11)\,10^{-17}\ {\rm s}\ .
\en

\section{Summary}
In this paper, we have reconsidered the chiral expansion
of the $\pi^0\to 2\gamma$ amplitude.
At first, we have focused on the two-flavour expansion. We have considered
the expansion beyond the known NLO (which we have expressed
in terms of the coupling constants introduced in ref.~\cite{girlanda}).
We have computed all the loop graphs which contribute at NNLO.
As expected, we found that the divergences are renormalizable by
Lagrangian terms of chiral order $p^8$ in the anomalous sector.
We found that chiral logarithms are present at this order. For physical values
of the quark masses $m_u$, $m_d$ these NNLO corrections turn out to be
negligible. Even the terms enhanced by logarithms are numerically very small
in practice. Our final expression (eq.~\rf{finalNNLO}) could be useful
in association with lattice QCD simulations in which unphysical quark masses
can be used.
This would provide a direct evaluation of the $SU(2)$ couplings.
As an interesting application, one could deduce (using
also experimental data such as from PrimEx) a precision determination
of $F_\pi$ uncorrelated with the value of $V_{ud}$.
Such simulations have
not yet been performed for correlation functions in the anomalous sector,
but this would be of obvious interest.

In order to perform a more detailed phenomenological analysis at present,
it is possible to enlarge the chiral expansion from $SU(2)$ to $SU(3)$.
This allows one to derive some information on the $SU(2)$ coupling
constants. We have derived the expansion of the $SU(2)$
couplings $\cwr{i}$ as a function of $m_s$ up to $O(m_s)$
and inserted this result into the $SU(2)$ expansion formula.
The leading, $1/m_s$ terms in this expansion, reflect the influence of
$\pi^0-\eta$ mixing.
We then implemented a modified chiral counting in which $m_s$ is counted
as $O(p)$ rather than $O(p^2)$.
This counting accommodates the fact that  $m_u$, $m_d$ are
significantly smaller than $m_s$. The formulas obtained in this way are
somewhat simpler and easier to interpret than those obtained in
the usual chiral
counting but the numerical results are essentially identical.

We have updated the inputs to be used in the chiral formula. A key input
is the value of $F_\pi$, the pion decay constant in pure QCD.
Another important input is the value of the
$\eta\to 2\gamma$ decay width, which we use to determine the value of the
$SU(3)$ LEC $\CW{8}$. In the chiral approach, this LEC encodes the
effect of $\eta-\eta'$ mixing. Our result agrees well with that of approaches
which account for $\eta-\eta'$ mixing explicitly,
using large $N_c$ arguments in addition to chiral
counting~\cite{kitazawa,kaisertempe,goityholstein}. The overall
uncertainty is dominated by the unknown terms of order $p^3$,
i.e. proportional to $m_u m_s$, $m_d m_s$ in the chiral expansion.
As a final remark, we note that $F_\pi$ is determined from the weak decay of
the $\pi^+$ assuming the validity of the standard model. Some recently
proposed Higgsless variants can accommodate deviations from the standard
$V-A$ coupling of quarks to the $W$ as large as a few percent~\cite{bernard}.
Precision measurements of the $\pi^0$ lifetime can provide constraints
on such models.

\medskip
\noindent{\Large\bf Acknowledgments}\\
We want to acknowledge useful comments and discussions with J.~Bijnens, G.~Ecker, B.~Jantzen, H.~Neufeld, R.~Rosenfelder and P. Talavera.
This work is
supported in part by the European commission MRTN FLAVIAnet [MRTN-CT-2006035482], Center for Particle Physics [LC 527] and
GACR [202/07/P249].
K.K. gratefully acknowledges the hospitality of the Institut de Physique Nucl\'eaire at Orsay during his visits.

\renewcommand{\theequation}{A-\arabic{equation}}
\setcounter{equation}{0}
\section*{Appendix I}
We give here the result of our computation of diagrams $(f)$ and $(g)$ in
fig.~\fig{2loopsdiags}:
\bea
&& F_\pi\, T_{(f)}= {M^4\over\pi^2 F^4}
\Bigg\{
-{11\over4}\left[ \Lambda^2
                 +\Lambda \left(L_\pi -{31\over1056\pi^2}\right)
+{1\over2} L_\pi^2  -{31\over1056\pi^2} L_\pi +{1\over6144\pi^2}\right]
\nonumber \\
&& -{467\over 98304 \pi^4} \Bigg\}\ ,\\
&& F_\pi\, T_{(g)}={M^4\over\pi^2 F^4}
\Bigg\{
{7\over3}\left[ \Lambda^2
               +\Lambda \left(L_\pi -{59\over1792\pi^2}\right)
+{1\over2} L_\pi^2  -{59\over1792\pi^2} L_\pi +{1\over6144\pi^2}\right] \nonumber \\
&& +{1\over512\pi^4}\left[ 3\sqrt3{\rm Cl_2}({\pi\over3})-{4\over3}\zeta(3)
-{1135\over576}
\right]
\Bigg\}\ ,
\ena
with
\be
\Lambda={1\over 16\pi^2 (d-4)}\ .
\en
\section*{Appendix II}
We collect below the expansions of the $SU(2)$ couplings $\cwr{i}$
as a function of $m_s$ up to $O(m_s)$. The notations $L_K$, $L_\eta$
and $c_0$ having been introduced in eqs.~\rf{c0} and \rf{Lk}
these expansions read
{\allowdisplaybreaks
\bea\lbl{ciwexp3}
&&\cwr{1} = \phantom{-2c_0} \CWr{2} - {1\over2} \CWr{3} +
{1\over4}{1\over(32\pi^2)^2 F_0^2} (  L_K+1+{1\over3} L_\eta )
\nonumber\\
&&\cwr{2} = \phantom{-2}c_0+ \CWr{4} - {1\over2} \CWr{5} +{3\over2} \CWr{6}
\nonumber\\
&&\cwr{3} = -{3\over2}c_0+\CWr{7} + 3 \CWr{8}
\nonumber\\
&&\cwr{4} = -{1\over2}c_0+ \CWr{9}+ 3 \CWr{10}
\nonumber\\
&&\cwr{5} = \phantom{-2c_0}\CWr{11} +
{1\over8}{1\over(32\pi^2)^2F_0^2}( L_K+1 + {2\over3} L_\eta)
\nonumber\\
&&\cwr{6} = \phantom{2}-c_0+ \CWr{5} - {3\over2} \CWr{6}
- {1\over2} \CWr{14} - {1\over2} \CWr{15}
\nonumber\\
&& \cwr{7} =\phantom{-}{3\over2}c_0 -3\CWr{8} + {1\over4} \CWr{22}
\nonumber\\
&& \cwr{8} =\phantom{-}{3\over4}c_0 + {1\over2} \CWr{7}+ 3 \CWr{8}
- {1\over8} \CWr{22}
\nonumber\\
&& \cwr{9} =\phantom{-2c_0}  - \CWr{13} +\CWr{14} + \CWr{15}
- {3\over2} {1\over(32\pi^2)^2F_0^2} (L_K+1)
\nonumber\\
&&\cwr{10} =\phantom{-2c_0}  \CWr{19}- \CWr{20}  - \CWr{21} -\CWr{22}
+ {3\over2} {1\over(32\pi^2)^2 F_0^2} (L_K+1)
\nonumber\\
&& \cwr{11} =\phantom{-2c_0} \CWr{22}
\nonumber\\
&& \cwr{12} = 0
\nonumber\\
&&\cwr{13} = \phantom{-2c_0}-2 \CWr{22} + {1\over(32\pi^2)^2F_0^2}(L_K+1)\ .
\ena
}


\begin{thebibliography}{99}
\bibitem{primex}
  M.~Kubantsev, I.~Larin and A.~Gasparyan  [PrimEx Collaboration],
  AIP Conf.\ Proc.\  {\bf 867} (2006) 51
  [arXiv:physics/0609201].

\bibitem{kitazawa}Y.~Kitazawa,
Phys.\ Lett.\ B {\bf 151} (1985) 165.


\bibitem{donoghuelin}
  J.~F.~Donoghue, B.~R.~Holstein and Y.~C.~R.~Lin,
  Phys.\ Rev.\ Lett.\  {\bf 55} (1985) 2766.

\bibitem{bijnenscornet}
  J.~Bijnens, A.~Bramon and F.~Cornet,
  Phys.\ Rev.\ Lett.\  {\bf 61} (1988) 1453.

\bibitem{riazuddin}  Riazuddin and Fayyazuddin,
  Phys.\ Rev.\  D {\bf 37} (1988) 149.

\bibitem{mou95}
B.~Moussallam,
Phys.\ Rev.\ D {\bf 51} (1995) 4939
[arXiv:hep-ph/9407402].

\bibitem{anantmou}
  B.~Ananthanarayan and B.~Moussallam,
  JHEP {\bf 0205} (2002) 052
  [arXiv:hep-ph/0205232].

\bibitem{nasrallah}
  N.~F.~Nasrallah,
  Phys.\ Rev.\  D {\bf 66} (2002) 076012.

\bibitem{kaisertempe}
  R.~Kaiser,
Proceedings of the Institute for Nuclear Theory- vol.12:
{\it Phenomenology of large $N_c$ QCD}, ed. R.F. Lebed,
World Scientific, Singapore (2002)
[http://www.slac.stanford.edu/spires/find/hep/www?irn=5533686]

\bibitem{goityholstein}  J.~L.~Goity, A.~M.~Bernstein and B.~R.~Holstein,
  Phys.\ Rev.\  D {\bf 66} (2002) 076014
  [arXiv:hep-ph/0206007].

\bibitem{ioffe}  B.~L.~Ioffe and A.~G.~Oganesian,
  Phys.\ Lett.\  B {\bf 647} (2007) 389
  [arXiv:hep-ph/0701077].


\bibitem{pagelszep}H.~Pagels and A.~Zepeda,
Phys.\ Rev.\ D {\bf 5} (1972) 3262.

\bibitem{scherrer02}  S.~Scherer,
  Adv.\ Nucl.\ Phys.\  {\bf 27} (2003) 277
  [arXiv:hep-ph/0210398].

\bibitem{urech95} R.~Urech,
  Nucl.\ Phys.\  B {\bf 433} (1995) 234
  [arXiv:hep-ph/9405341].

\bibitem{cohen08}
  S.~D.~Cohen, H.~W.~Lin, J.~Dudek and R.~G.~Edwards,
  arXiv:0810.5550 [hep-lat].

\bibitem{colangelo1}
 G.~Colangelo,
  Phys.\ Lett.\  B {\bf 350} (1995) 85
  [Erratum-ibid.\  B {\bf 361} (1995) 234]
  [arXiv:hep-ph/9502285].

\bibitem{colangelo2}
  J.~Bijnens, G.~Colangelo and G.~Ecker,
  Phys.\ Lett.\  B {\bf 441} (1998) 437
  [arXiv:hep-ph/9808421].

\bibitem{weinberg79}
  S.~Weinberg,
  Physica A {\bf 96} (1979) 327.

\bibitem{WZ}J.~Wess and B.~Zumino,
Phys.\ Lett.\ B {\bf 37} (1971) 95,
E.~Witten,
Nucl.\ Phys.\ B {\bf 223} (1983) 422,
R.~Kaiser,
Phys.\ Rev.\ D {\bf 63} (2001) 076010
[arXiv:hep-ph/0011377].


\bibitem{ABJ}
S.~L.~Adler,
Phys.\ Rev.\  {\bf 177} (1969) 2426,
J.~S.~Bell and R.~Jackiw,
Nuovo Cim.\ A {\bf 60} (1969) 47,
W.~A.~Bardeen,
Phys.\ Rev.\  {\bf 184} (1969) 1848.


\bibitem{fearing}
  H.~W.~Fearing and S.~Scherer,
  Phys.\ Rev.\  D {\bf 53} (1996) 315
  [arXiv:hep-ph/9408346].

\bibitem{akhoury}
R.~Akhoury and A.~Alfakih,
Annals Phys.\  {\bf 210} (1991) 81.

\bibitem{girlanda}
  J.~Bijnens, L.~Girlanda and P.~Talavera,
  Eur.\ Phys.\ J.\  C {\bf 23} (2002) 539
  [arXiv:hep-ph/0110400].

\bibitem{eberthauser}
  T.~Ebertshauser, H.~W.~Fearing and S.~Scherer,
  Phys.\ Rev.\  D {\bf 65}, 054033 (2002)
  [arXiv:hep-ph/0110261].

\bibitem{gl84}
J.~Gasser and H.~Leutwyler,
Annals Phys.\  {\bf 158} (1984) 142.

\bibitem{karolfirst}
  K.~Kampf and J.~Novotny,
  Acta Phys.\ Slov.\  {\bf 52} (2002) 265
  [arXiv:hep-ph/0210074].

\bibitem{burgi}
  U.~B\"urgi,
  Nucl.\ Phys.\  B {\bf 479} (1996) 392
  [arXiv:hep-ph/9602429].

\bibitem{bijetalpipi2}
  J.~Bijnens, G.~Colangelo, G.~Ecker, J.~Gasser and M.~E.~Sainio,
  Nucl.\ Phys.\  B {\bf 508} (1997) 263
  [Erratum-ibid.\  B {\bf 517} (1998) 639]
  [arXiv:hep-ph/9707291].

\bibitem{bce99a}
  J.~Bijnens, G.~Colangelo and G.~Ecker,
  JHEP {\bf 9902} (1999) 020
  [arXiv:hep-ph/9902437].

\bibitem{gassersainio}
 J.~Gasser and M.~E.~Sainio,
  Eur.\ Phys.\ J.\  C {\bf 6} (1999) 297
  [arXiv:hep-ph/9803251].


\bibitem{gl85}
J.~Gasser and H.~Leutwyler,
Nucl.\ Phys.\ B {\bf 250} (1985) 465.

\bibitem{gasserhaefeli}
  J.~Gasser, C.~Haefeli, M.~A.~Ivanov and M.~Schmid,
  Phys.\ Lett.\  B {\bf 652} (2007) 21
  [arXiv:0706.0955 [hep-ph]].

\bibitem{haefeli}
  C.~Haefeli, M.~A.~Ivanov and M.~Schmid,
  Eur.\ Phys.\ J.\  C {\bf 53} (2008) 549
  [arXiv:0710.5432 [hep-ph]].


\bibitem{marcianosir}
W.~J.~Marciano and A.~Sirlin,
Phys.\ Rev.\ Lett.\  {\bf 71} (1993) 3629.

\bibitem{neufeldknecht}
M.~Knecht, H.~Neufeld, H.~Rupertsberger and P.~Talavera,
Eur.\ Phys.\ J.\ C {\bf 12} (2000) 469
[arXiv:hep-ph/9909284].

\bibitem{descotesmou}
  S.~Descotes-Genon and B.~Moussallam,
  Eur.\ Phys.\ J.\  C {\bf 42} (2005) 403
  [arXiv:hep-ph/0505077].

\bibitem{townerhardy}
  I.~S.~Towner and J.~C.~Hardy,
  Phys.\ Rev.\  C {\bf 77} (2008) 025501
  [arXiv:0710.3181 [nucl-th]].

\bibitem{kaplanman}
  D.~B.~Kaplan and A.~V.~Manohar,
  Phys.\ Rev.\ Lett.\  {\bf 56} (1986) 2004.

\bibitem{allton2+1}
  C.~Allton {\it et al.}  [RBC-UKQCD Collaboration],
  arXiv:0804.0473 [hep-lat].


\bibitem{bijnensghorbani}  J.~Bijnens and K.~Ghorbani,
  JHEP {\bf 0711} (2007) 030
  [arXiv:0709.0230 [hep-ph]].

\bibitem{kamboreta3piem}
  R.~Baur, J.~Kambor and D.~Wyler,
  Nucl.\ Phys.\  B {\bf 460} (1996) 127
  [arXiv:hep-ph/9510396].

\bibitem{kubiseta3piem}
  C.~Ditsche, B.~Kubis and U.~G.~Meissner,
  arXiv:0812.0344 [hep-ph].

\bibitem{Kastner-Neufeld}
A.~Kastner and H.~Neufeld,
  arXiv:0805.2222 [hep-ph].


\bibitem{karoletmoi}
  K.~Kampf and B.~Moussallam,
  Eur.\ Phys.\ J.\  C {\bf 47} (2006) 723
  [arXiv:hep-ph/0604125].

\bibitem{deandrea}
  A.~Deandrea, A.~Nehme and P.~Talavera,
  Phys.\ Rev.\  D {\bf 78} (2008) 034032

\bibitem{bellepiprime}
  K.~Abe {\it et al.}  [Belle Collaboration],
  arXiv:hep-ex/0610022.

\bibitem{pham}
  T.~N.~Pham,
  Phys.\ Lett.\  B {\bf 246} (1990) 175.

\bibitem{terning}
  B.~Holdom, J.~Terning and K.~Verbeek,
  Phys.\ Lett.\  B {\bf 245} (1990) 612.

\bibitem{bijprad94}
  J.~Bijnens and J.~Prades,
  Z.\ Phys.\  C {\bf 64} (1994) 475
  [arXiv:hep-ph/9403233].

\bibitem{moussall95}
  B.~Moussallam,
  Phys.\ Rev.\  D {\bf 51} (1995) 4939
  [arXiv:hep-ph/9407402].

\bibitem{browman}  A.~Browman, J.~DeWire, B.~Gittelman, K.~M.~Hanson, E.~Loh and R.~Lewis,
  Phys.\ Rev.\ Lett.\  {\bf 32} (1974) 1067.

\bibitem{rodrigues}
  T.~E.~Rodrigues {\it et al.},
  Phys.\ Rev.\ Lett.\  {\bf 101} (2008) 012301.

\bibitem{pdg}
  C.~Amsler {\it et al.}  [Particle Data Group],
  Phys.\ Lett.\  B {\bf 667} (2008) 1.

\bibitem{antonelli}  M.~Antonelli,
  arXiv:0712.0734 [hep-ex].

\bibitem{dalitz}
  K.~Kampf, M.~Knecht and J.~Novotny,
  Eur.\ Phys.\ J.\  C {\bf 46} (2006) 191
  [arXiv:hep-ph/0510021].

\bibitem{bernard}  V.~Bernard, M.~Oertel, E.~Passemar and J.~Stern,
  JHEP {\bf 0801} (2008) 015
  [arXiv:0707.4194 [hep-ph]].


\end{thebibliography}
\end{document}